\documentclass{article}

\usepackage{PRIMEarxiv}

\usepackage[utf8]{inputenc} 
\usepackage[T1]{fontenc}    
\usepackage{hyperref}       
\usepackage{url}            
\usepackage{booktabs}       
\usepackage{amsfonts}       
\usepackage{nicefrac}       
\usepackage{microtype}      
\usepackage{lipsum}
\usepackage{fancyhdr}       
\usepackage{graphicx}       
\graphicspath{{media/}}     

\title{A Novel Three-Dimensional Navigation Method for the Visually Impaired}

\author{
  Stanley Shen \\
  Los Gatos High School \\
  Los Gatos\\
  {stanleyshenlg@gmail.com} \\
}

\begin{document}
\maketitle

\begin{abstract}
According to the World Health Organization \cite{WHO}, visual impairment is estimated to affect
approximately 2.2 billion people worldwide. The visually impaired must currently rely on
navigational aids to replace their sense of sight, like a white cane or GPS (Global Positioning
System) based navigation, both of which fail to work well indoors. The white cane cannot be
used to determine a user’s position within a room, while GPS can often lose connection indoors
and does not provide orientation information, making both approaches unsuitable for indoor use. Therefore, this research seeks to develop a 3D-imaging solution that enables contactless navigation through a complex indoor environment. The device can pinpoint a user’s position and
orientation with 31\% less error compared to previous approaches while requiring only 53.1\% of
the memory, and processing 125\% faster. The device can also detect obstacles with 60.2\% more
accuracy than the previous state-of-the-art models while requiring only 41\% of the memory and
processing 260\% faster. When testing with human participants, the device allows for a 94.5\%
reduction in collisions with obstacles in the environment and allows for a 48.3\% increase in
walking speed, showing that my device enables safer and more rapid navigation for the visually
impaired. All in all, this research demonstrates a 3D-based navigation system for the visually
impaired. The approach can be used by a wide variety of mobile low-power devices, like cell
phones, ensuring this research remains accessible to all.
\end{abstract}

\section{Introduction}

\indent Visual impairment, the deterioration of the ability to see, is estimated to affect approximately 285 million people worldwide. Of those 285 million, 39 million are completely blind and 246 million have poor vision. \cite{WHO}. Some possible causes are diabetic retinopathy and glaucoma. Due to their poor eyesight, the blind must rely on their ability to non-visually memorize their surroundings and their obstacles. This makes it difficult for them to navigate new environments, especially those less forgiving to trial and error in high foot-traffic areas. It can even be hazardous for other areas, like crosswalk intersections, where the risk of getting hit by a car is high, or a storage room with hard and heavy objects. Even though the world has taken efforts to make society an easier place to navigate, such as installing Accessible Pedestrian Signals (APS) systems at crosswalks (which beep or chirp when it is safe to cross), blind pedestrians still note that sounds from the speakers did not provide adequate information, resulting in 97\% of physical therapists reporting that their visually impaired students could not effectively line up to cross a street \cite{Carter2015}. This pattern continues in indoor environments, where the visually impaired have even fewer accommodations. Sometimes, the only accommodation that exists is tiny braille markings indicating room numbers on doors or walls, which does not assist the visually impaired in finding the room in the first place. Other places might not have any markings at all. For example, in a survey conducted about issues faced when walking around indoors, the majority of visually impaired "often" or "sometimes" experience issues with ground level, body level, and head level obstacles \cite{Jeamwatthanachai2019}. Currently, commonly employed solutions by the visually impaired either involve a white cane or a human guide.  White canes are not ideal because they cannot detect any obstacles that are not resting on the ground, such as low-hanging tree branches. Another option would be sighted guides, as all human environments can be easily navigated with the help of a sighted guide, but having a human guide is not always possible as sighted guides might not be available to help, and sighted guides can restrict personal independence. Guides are not ideal because it requires the presence of a sighted chaperone to give instructions. Other solutions, like guided dogs, also are not optimal as they are often expensive costing up to 37,000 USD, need to be trained (which is a long and time-consuming process), and are not allowed in all locations, such as taxis \cite{Jeamwatthanachai2019}. Current technological solutions use Global Positioning Systems (GPS) and Geographic Information Systems (GIS) to give directions for the users to walk from one location to another. These solutions are unable to detect obstacles not mapped by GPS and satellite imagery, so these solutions cannot be used for day-to-day use in the real world without the use of an additional obstacle-avoidance system \cite{Kuriakose2020} 

The goal of this project was to create an end-to-end (complete) technological navigation aid for the visually impaired. The research aims to develop a machine learning model to perform camera pose estimation or camera localization based on a standard camera phone sensor to avoid the need for bulky 3D sensors. Then, this research also worked on development of a lightweight 3D object-detection algorithm based on existing state-of-the-art (SOTA) algorithms, to enable people without powerful computers to also run the developed software. Finally, the research will integrate both algorithms and use them to power a voice interface to communicate navigation information to the user, such as the position of objects of interest, or the position of objects to avoid.

\subsection{Visually Impaired Navigation}

For centuries, the visually impaired have had to rely on low-tech options for navigation, such as the white cane. This is a slow and inefficient method that is also quite risky and hard to use in day-to-day life. Canes have not fundamentally changed in function since their inception, except being manufactured out of lower-weight plastics or composite materials or having folding hinges to allow for navigation in smaller spaces, neither of which significantly improve navigation accuracy and speed. 
However, recently, some researchers have begun integrating computer technology into smart canes to make them easier and faster to use. One prominent example is the cane designed by Slade et al. While groundbreaking and much more affordable compared to previous electronically-guided canes (the design cost only 400 USD to make, compared to 5,000 USD of previous designs) \cite{Slade2021}, the cane still primarily makes use of expensive, bulky 2D LiDAR sensors (sensors with an array of lasers which measure distance), which have to be carried around. Users complained about the weight of the device as a recurring usability issue. Therefore, the goal of this research is to remove the need for a 3D scanner for daily use. The device first creates a baseline scan of the environment. Since the developed method will use true 3D scans of the surrounding environment, rather than a flat planar 2D “bird’s eye” view, the device can also use object detection to find the size, position, and description of each object. This enables users to search for and locate an object, making the method more useful when indoors. Also, instead of using a 3D scanner during daily operation, the device uses a camera to localize the user (find the location of everything else relative to the camera). This means the project can be broken into three main sections: 1) camera pose localization, 2) object detection, and 3) system integration into a unified device for the end-user.
\subsection{Camera Pose Localization}
Camera Pose Localization, which is the task of inferring a camera’s position, has been a long-standing challenge in the field of robotics. This is the first and crucial part of this research, as the model will need the location of the user to provide accurate and effective navigational instructions. If the position is wrong, the device might give incorrect location directions to the user. The traditional approach is to use a simultaneous localization and mapping (SLAM) approach; however, this is computationally intensive and requires the use of high-resolution and high-frequency sensors. 

This is an issue because these sensors require large batteries to supply power to the sensor, as well as powerful computers to process the data. This is because the approach uses frame to frame odometry, using movement between each frame from a 3D sensor to “match” the perspective of the sensor and find the position of the camera. However, frame to frame odometry requires frame to frame correspondence, meaning that movement needs to be matched to the previous frame before it is needed, or else the SLAM system “loses track” of its position. 

SLAM also cannot relocalize (determine initial position) the 3D-based absolute position, as the position is determined by frame-to-frame differences. The requirement of tracking movement frame-to-frame means that a high sample rate is required for high accuracy, making SLAM an unsuitable system for deployment in low-power mobile systems. 

However, other systems for localization do exist, such as appearance-based localization methods. These methods can determine the camera’s position using details from the environment. Appearance-based localization “matches” the current perspective from a camera to a database of images with known positions, and uses the most similar image from a database to make a prediction. These methods are only capable of rough localization, which is not accurate enough for navigating tight spaces. As a result, researchers began to develop methods that use machine learning, which is what this research attempts to build on.

The first major approach was the appearance-based method PoseNet, a machine learning approach, described and implemented in the paper PoseNet: A Convolutional Network for Real-Time 6-DOF Camera Relocalization  \cite{Kendall2015}. This method uses a general backbone “feature extraction” neural network (artificial collection of computer-simulated neurons) to extract features of the images, similar to the SIFT (Scale-invariant feature transform)-based approaches for localization. In the research paper, an outdated GoogLeNet \cite{Szegedy2014}(aka Inception V1) is used as the backbone. The second section is the “pose regression” neural network, which then converts the features of each image into position and orientation information. This network, while groundbreaking at the paper’s release, is largely hampered by the use of a now older feature extraction backbone, and replacing it will likely vastly improve the localization performance of the neural network. This is important as navigation requires precise knowledge of a person’s location in a room, meaning improving the performance is important. 

Another method is to use kinematics equations to determine the position from the linear and angular acceleration sensors derived from accelerometers and gyroscopes. However, due to the double integration needed to convert acceleration to position, this leads to a large amount of noise, error, and jitter (rapid changes in output estimation) in the resulting position and orientation of the camera. Even applying filters, such as Mahony, Madgwick, or Kalman filters, which should help reduce noise, leads to large amounts of drift (deviation) in the resulting position, since small movements are filtered out to reduce noise. This means that dead reckoning, or determining position and orientation by considering velocity and changes in orientation over time, is a difficult task to perform accurately. This research adapts this method to run in real-time by correcting the drift immediately after each step, instead of running the drift compensation “offline” (not real-time), which will allow use in this final device.

\subsection{Object Detection}

3D Data, and as a result, point clouds, have many benefits. For instinct, point clouds allow for route planning throughout a room, allowing for more efficient travel. CNNs, with convolution filters, run easily with images as images can be easily converted into a “grid” of pixels. However, point clouds are unevenly spaced. While pixels are taken based on the pixels in an image sensor, which are always fixed evenly, the density of a point cloud can vary greatly due to the design of the 3D sensors. Point clouds are also invariant, as re-ordering the order of the points (when listed in an array-like data structure) leads to the same object, a property that does not apply to other types of data, like images. This makes 3D objects a difficult medium to apply machine learning directly on. Working with point clouds (3D data), even without the context of object detection, has long been a hard issue to solve. 

\subsection{System Integration}
	
The last section is integration, which is combining pose localization and object detection into one singular device and app. Slade’s cane \cite{Slade2021}, which was designed primarily for outdoor navigation, was a white cane with a motor attached to the end to provide force feedback regarding which direction to travel to avoid a collision. However, given that this device will be used indoors, the research developed and demonstrated a completely new interface that will also give information regarding the relative height of obstacles. 

\section{Methods}

My research can be divided into three main phases: camera pose localization, object detection, and system integration into a unified device for the user.

\subsection{Camera Pose Localization}
As stated in the methods, this research used PoseNet \cite{Kendall2015} as a base neural network. This research focused on augmenting the performance of the feature extraction section (which isolates important details within a CNN). However, by providing more accurate and detailed features of the image, the pose regressor should be able to produce more accurate camera pose information. This is because more accurate and important details of the room help neural networks create accurate predictions. The original work by Kendall et al. used GoogLeNet, a CNN, with 22 layers extracting different features \cite{Szegedy2014}. However, more advanced techniques have been introduced for image classification, which can be used as the basis for a feature extraction system that can be modified for pose regression. One example would be Residual Learning, which helps train deeper neural networks with more layers. Deep neural networks with large numbers of layers can multiply out the input values to have incredibly small gradients, leading to slow or no training. By allowing extracted details of the room to skip layers, this issue is partly reduced as direct skips discourage gradients from becoming too small.
The second method applied is fully connected CNN layers. This method, as referenced in the DenseNet paper \cite{Huang2016}, helps with alleviating the vanishing gradient problem. It also makes the network more efficient by reusing features, as well as reducing memory by reducing the number of required parameters. By connecting all the layers, all of the different layers “work together” to extract more features from an image. However, training to achieve efficient feature sharing while maintaining accuracy requires a large amount of training data. 
The third method is to use transformers (a network based on self-attention). Self-attention helps extract features, but the original approach of directly applying images to the network would require each pixel in the images to be compared to all the other pixels in the images, resulting in a quadratic increase in computation resources needed as image size increases. Given that images require significantly more memory than text, a vanilla transformer architecture becomes infeasible for image resolution beyond a few pixels. This is because each pixel is compared to every single other pixel. To circumvent this issue, this research used the technique of image patching or splitting the image into smaller patches, as first described in An Image is Worth 16x16 Words: Transformers for Image Recognition at Scale \cite{Dosovitskiy2020}. This way, the computational requirement can be reduced, as well as memory requirements, and take advantage of the ability of self-attention, which is used to find correlations between parts of an image, compared to CNNs, which only work on the section being processed by the filter. Lower computational requirements are paramount since they allow the device to give instructions more frequently, allowing the user to navigate more rapidly. 
To conduct dead reckoning through sensors, devices contain an IMU (inertial measurement unit) that can be positioned near a shoe.  This resets the integration error every single time the foot hits the ground. Dead reckoning is determining the position and orientation tracking without outside references. This is called gait tracking, as the device tracks the position of the foot. After every step, the foot is momentarily still before the person resumes the step, meaning the algorithm can set the calculated velocity to zero, with the purpose of minimizing the calculated drift in the process. The device also fused the position and orientation value from the machine learning algorithm and sensors, which helped create more stable and accurate results. 

\subsection{Object Detection}
One way that this research applied machine learning to 3D objects is by splitting the 3D object shape into a 3D grid in a process called voxelization. This way, the study could expand standard CNN filters into the third dimension. However, due to adding another dimension, computation and memory requirements increase exponentially. So, spare CNN  filters from Sparse 3D convolutional neural networks\cite{Graham15}are used. They work with a precomputed hash map with all the occupied areas, and skip processing whatever is in empty spaces, reducing memory and computation time. Compared to more complicated algorithms that have been applied to 3D objects, like transformers, CNNs can easily scale to large and smaller scenes. It does this by applying the same filter over, as well as reducing memory usage, compared to the more memory-intensive transformers, whose memory requirement increases quadratically as the number of points in a 3D scene doubles.

\subsection{System Integration}
One way that this research applied machine learning to 3D objects is by splitting the 3D object shape into a 3D grid in a process called voxelization. This way, the study could expand standard CNN filters into the third dimension. However, due to adding another dimension, computation and memory requirements increase exponentially. So, spare CNN  filters from Sparse 3D convolutional neural networks\cite{Graham15} are used. They work with a precomputed hash map with all the occupied areas, and skip processing whatever is in empty spaces, reducing memory and computation time. Compared to more complicated algorithms that have been applied to 3D objects, like transformers, CNNs can easily scale to large and smaller scenes. It does this by applying the same filter over, as well as reducing memory usage, compared to the more memory-intensive transformers, whose memory requirement increases quadratically as the number of points in a 3D scene doubles.

\section{Results}

\subsection{Dead Reckoning Based Pose Localization}

\begin{table}[h] 
\begin{tabular}{ |c|c| } 
 \hline
 Type of Position Tracking & Deviation Amount from Offline Reconstruction (Meters)  \\ 
 \hline
 Position Tracking, no Gait Compensation & 340.98 \\ 
 Offline Position Tracking, with Gait Compensation & 0 \\ 
 Online Position Tracking, with Gait Compensation & 0 \\
 \hline
\end{tabular}
\caption{Comparison of Dead Reckoning Methods} 
\label{tab:table}
\end{table}

\begin{figure}[!htb]
\minipage{0.32\textwidth}
  \includegraphics[width=\linewidth]{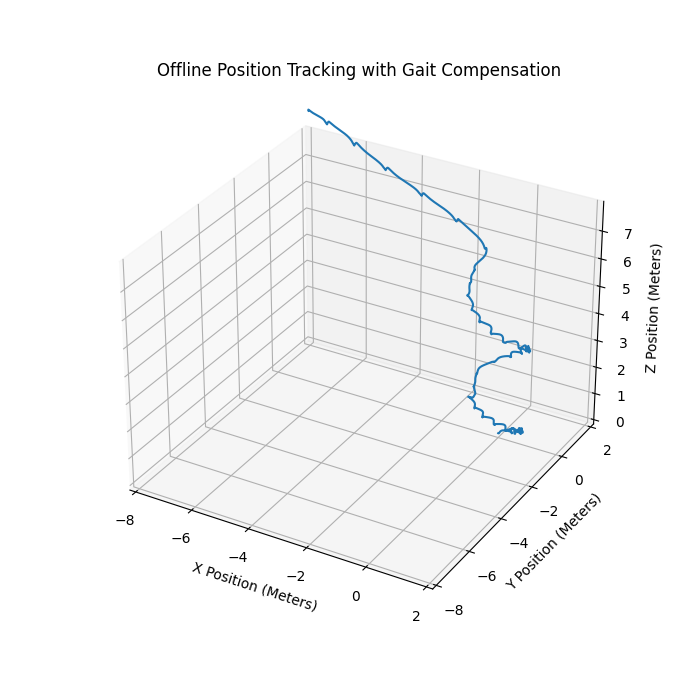}
  \caption{Estimated Position Graph}\label{fig:figure1}
\endminipage\hfill
\minipage{0.32\textwidth}
  \includegraphics[width=\linewidth]{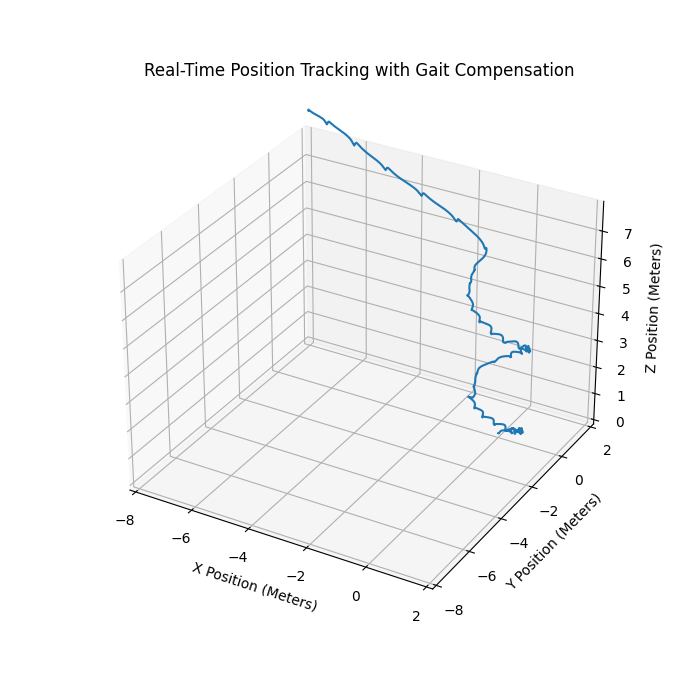}
  \caption{Estimated Position Graph}\label{fig:figure2}
\endminipage\hfill
\minipage{0.32\textwidth}%
  \includegraphics[width=\linewidth]{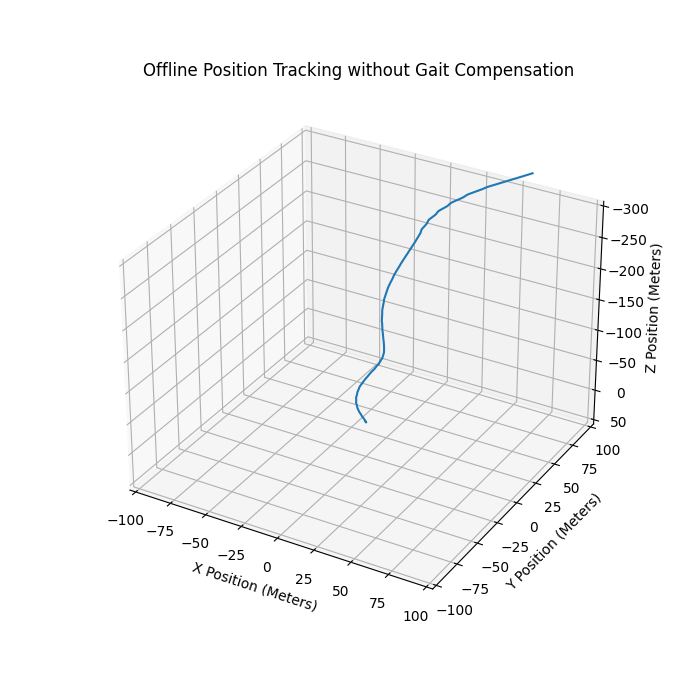}
  \caption{Estimated Position Graph}\label{fig:figure3}
\endminipage
\end{figure}

As seen in the 3D graphs of someone climbing up a spiral staircase and walking down a straight corridor, the offline position tracking without gait compensation (Fig 3), the results of the position traces are drifting far beyond recognition. However, compared to offline position tracking developed by Seb Madgwick (Fig 1), to the developed real-time adoption of his method (Fig 2), which runs in near-real-time (processing everything time a step is completed). The table and 3D graphs form much more recognizable trajectories of the spiral staircase, with both results being credibly similar or identical trajectories. 

\subsection{Machine Learning-based Pose Localization}

\begin{figure}[!htb]
\begin{center}
\minipage{0.6\textwidth}
  \includegraphics[width=\linewidth]{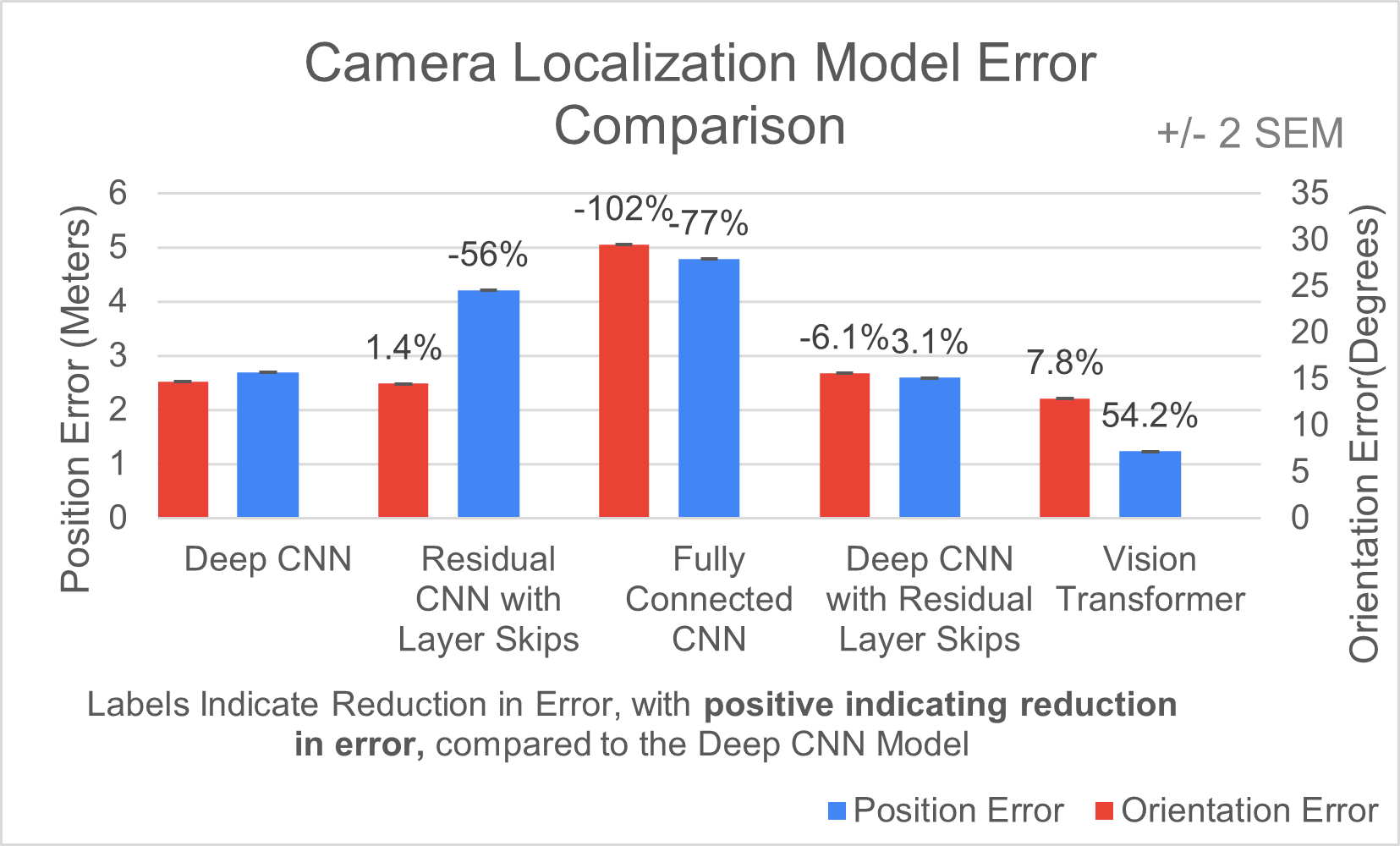}
  \caption{Error of Localization Model}\label{fig:figure4}
\endminipage
\end{center}
\end{figure}

\begin{figure}[!htb]
\begin{center}
\minipage{0.6\textwidth}
  \includegraphics[width=\linewidth]{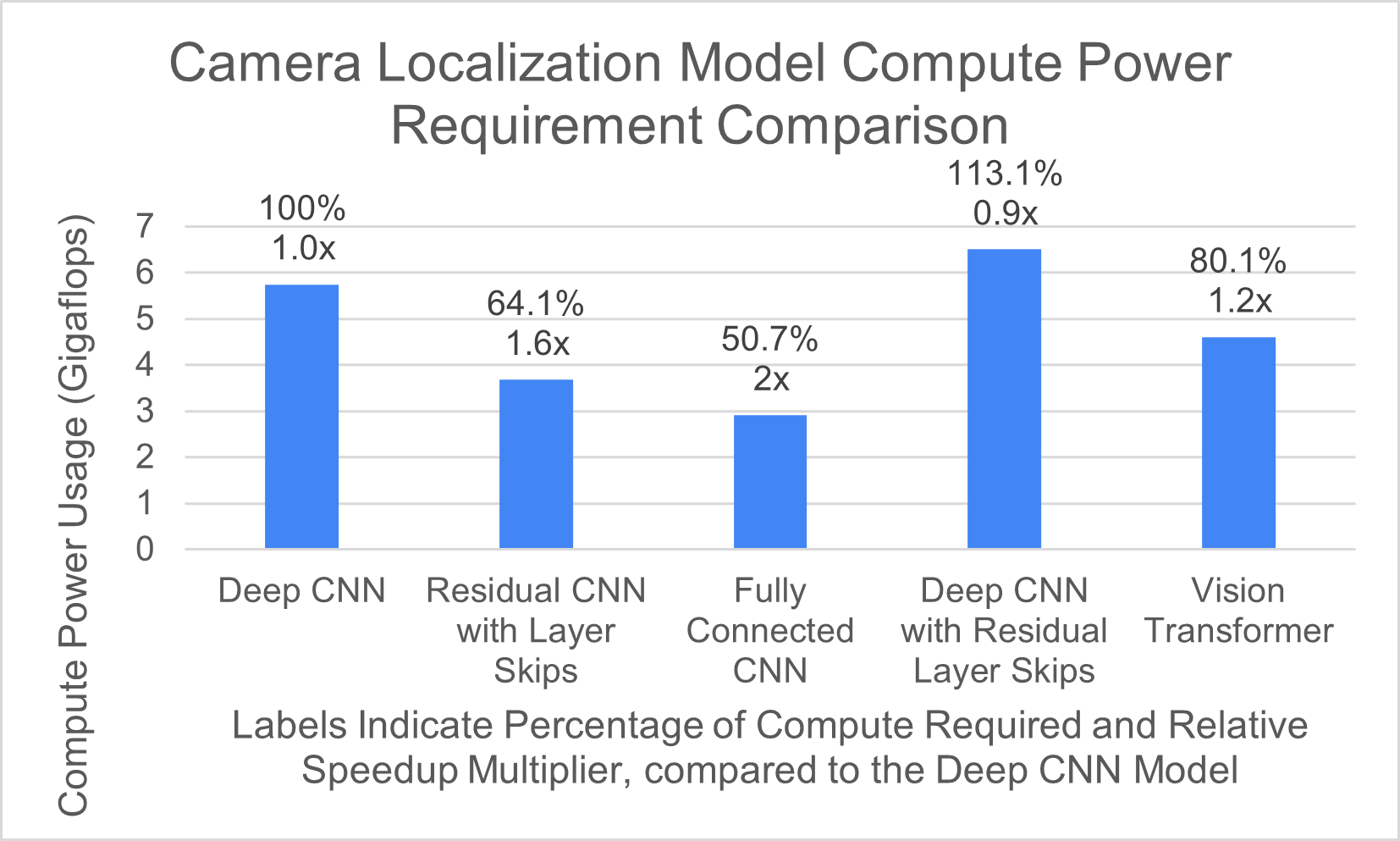}
  \caption{Compute power required for Localization Model}\label{fig:figure5}
\endminipage
\end{center}
\end{figure}

\begin{figure}[!htb]
\begin{center}
\minipage{0.6\textwidth}
  \includegraphics[width=\linewidth]{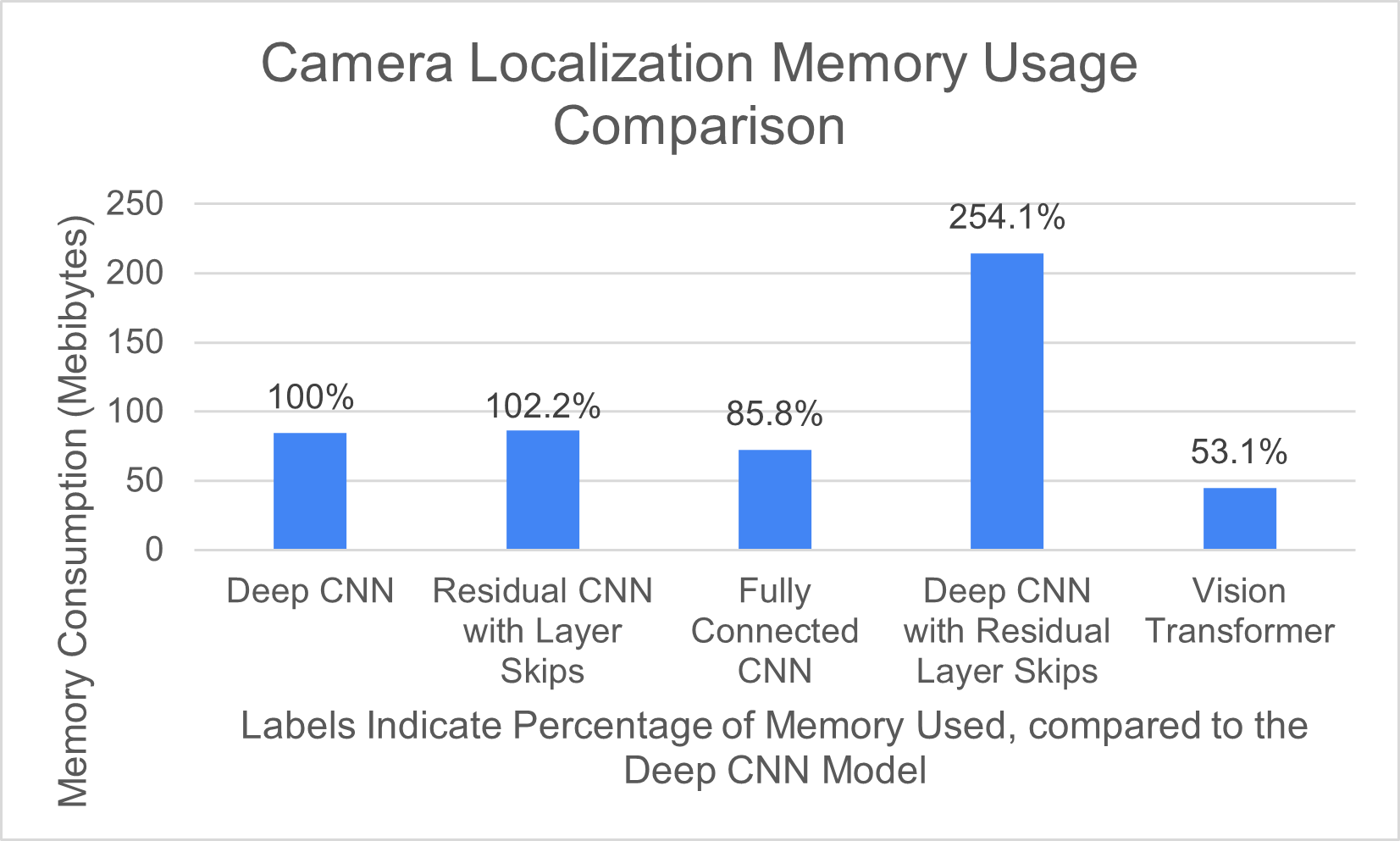}
  \caption{Memory required for Localization Model}\label{fig:figure6}
\endminipage
\end{center}
\end{figure}

To analyze the results, the accuracy (in the form of Position and Orientation Error in Fig 4), as well as computation and memory requirements (in the form of parameters per model and gigaflops per image in Fig 5 and Fig 6) are graphed out. It is important to have high accuracy to ensure precise location data, which makes navigation much more stable and safer. It is also important to have low parameters per model to reduce the space required to store and run a model as well as a low number of FLOPS(floating point operations per second) needed to execute a model to reduce the need for expensive power-intensive hardware. 

As shown in the graph, vision transformers perform with the highest accuracy in both positions and also have the average computation and memory requirements. Other methods, such as Fully Connected CNNs and Residual CNNs, due have lower processing requirements but offer worse performance compared to vision transformers. All in all, by switching from deep CNNs to vision transformers, the developed network can reduce error on average 31\% while processing data 1.2x faster and requiring 53.1

\subsection{Object Detection}

For the object detection neural network, this study mainly focused on the balance of the number of “levels” to encode. In theory, increasing the number of levels and increasing the depth should mean more features extracted from the 3D object, and as a result, better object detection, but this comes at the cost of requiring more memory and computation power. The memory needed in megabytes of RAM and the computation power needed were evaluated by measuring how many batches can be processed per second, with more batch per second meaning that less computational power is required. However, sometimes the computer cannot even run one batch on a GTX 1060 3GB GPU, so some configurations are listed with OOM (out of memory), meaning the GPU ran out of memory. 

\begin{figure}[!htb]
\begin{center}
\minipage{0.6\textwidth}
  \includegraphics[width=\linewidth]{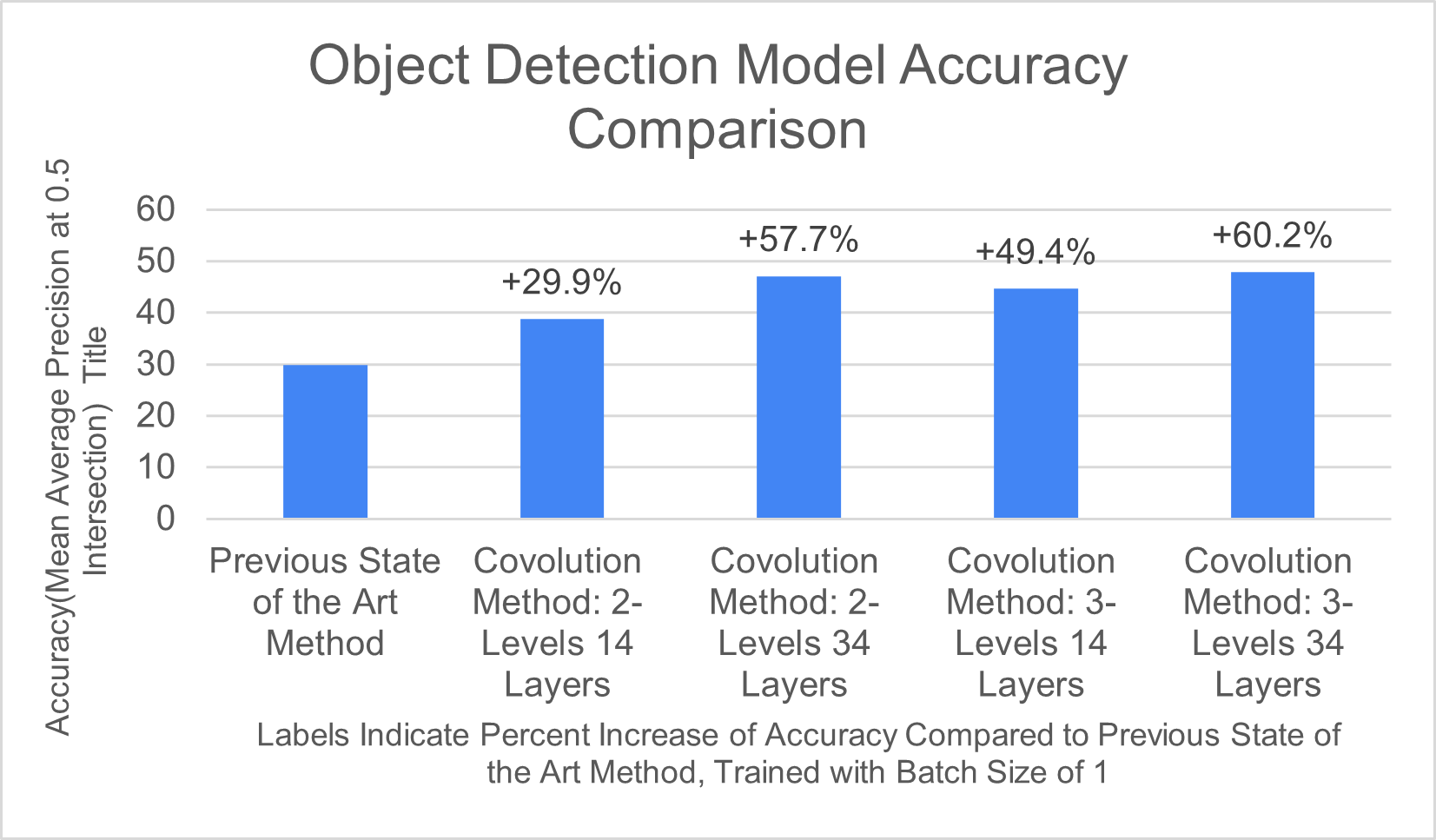}
  \caption{Accuracy of Object Detection Model}\label{fig:figure7}
\endminipage
\end{center}
\end{figure}

\begin{figure}[!htb]
\begin{center}
\minipage{0.6\textwidth}
  \includegraphics[width=\linewidth]{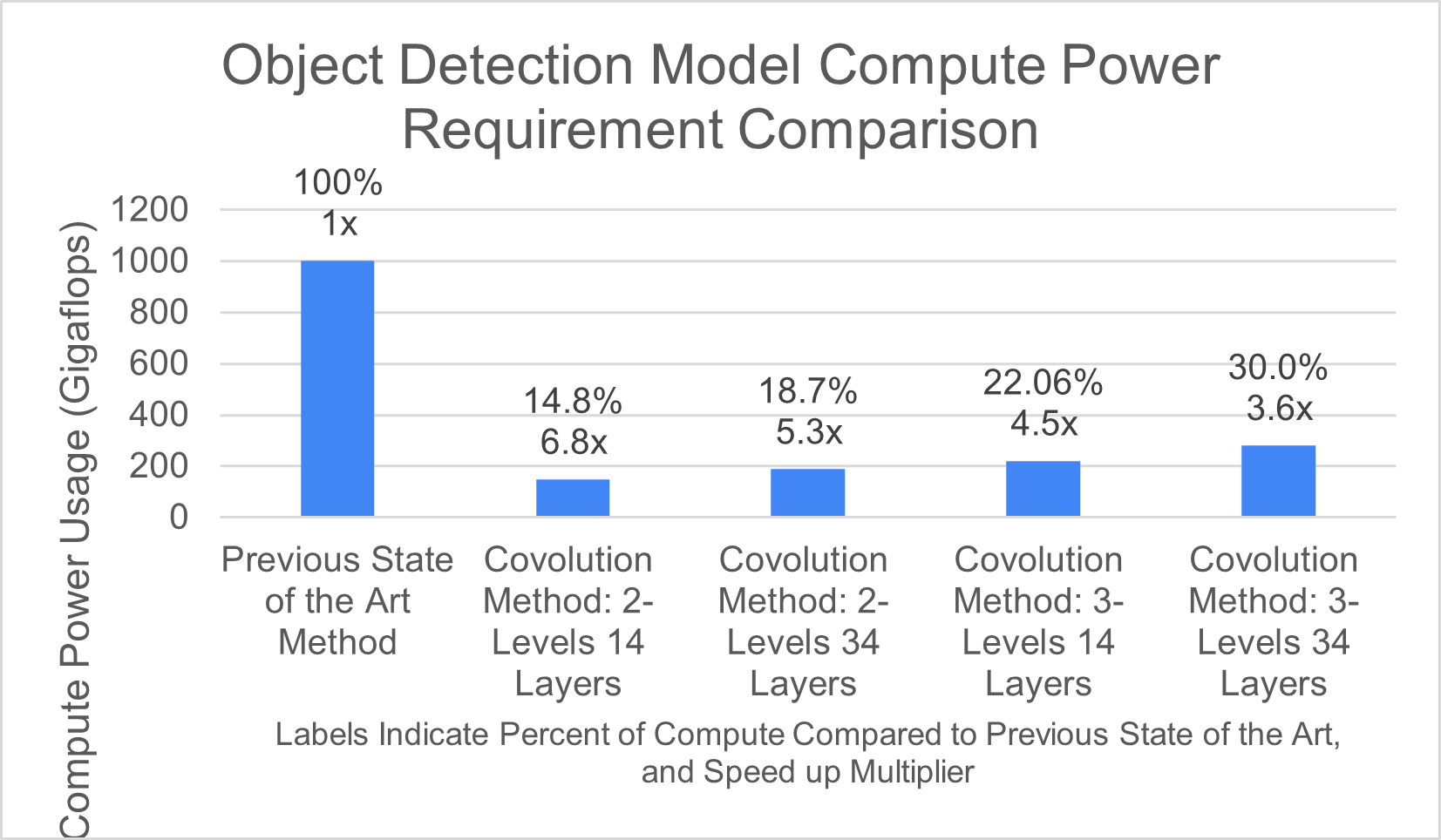}
  \caption{Compute Power of Object Detection Model}\label{fig:figure8}
\endminipage
\end{center}
\end{figure}

\begin{figure}[!htb]
\begin{center}
\minipage{0.6\textwidth}
  \includegraphics[width=\linewidth]{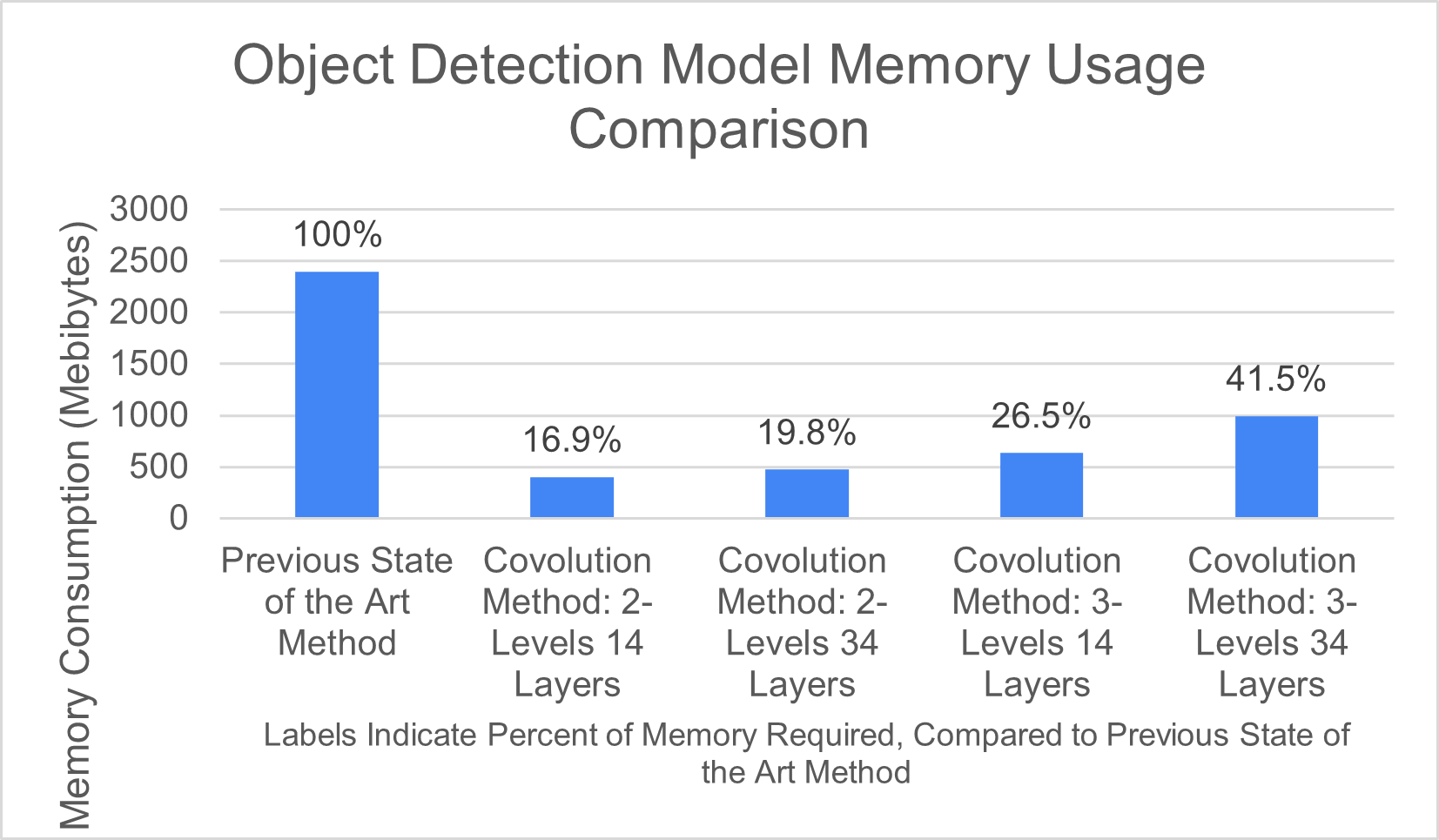}
  \caption{Memory required for Object Detection Model}\label{fig:figure9}
\endminipage
\end{center}
\end{figure}

\begin{figure}[!htb]
\begin{center}
\minipage{0.6\textwidth}
  \includegraphics[width=\linewidth]{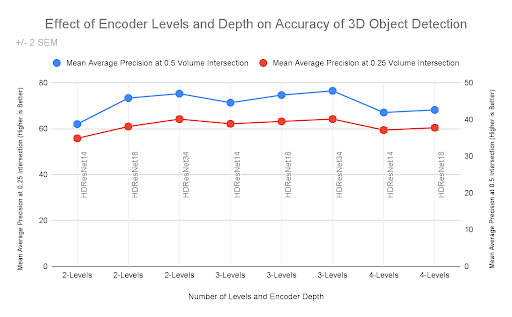}
  \caption{Memory required for Localization Model}\label{fig:figure10}
\endminipage
\end{center}
\end{figure}

\begin{figure}[!htb]
\begin{center}
\minipage{0.6\textwidth}
  \includegraphics[width=\linewidth]{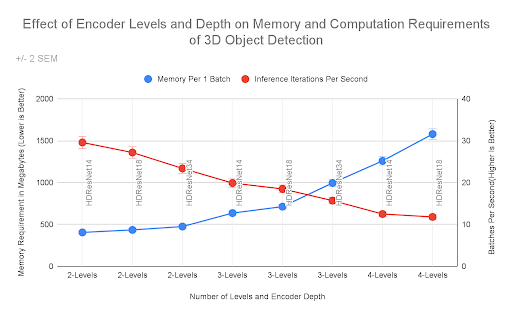}
  \caption{Memory required for Localization Model}\label{fig:figure11}
\endminipage
\end{center}
\end{figure}

As shown in Figure 11, increasing the number of levels and using a deeper pre-encoder (denoted by the number after the prefix HDResNet) rapidly increases the amount of memory required and leads to more computation power being required, as the number of batches (3D scenes) that can be processed per second, goes down rapidly. Fig 10 shows that increasing the number of levels does not lead to ever-increases of accuracy, and at around 3 levels, the accuracy reaches the highest point.

\subsection{Human Testing}
\newpage

In order to determine how much these performance and speed improvements would carry over to the real world, human trials were carried out. In order to ensure fair testing, participants were blindfolded with opaque lab goggles, moved to a random point in the room, and told to navigate a certain point within the room. Then, collisions with the environment and time to complete transit were recorded. The same trial was run again with randomized points, but this time, the participants used the developed device to navigate, instead of relying on their hands to walk around.

\begin{figure}[!htb]
\begin{center}
\minipage{0.6\textwidth}
  \includegraphics[width=\linewidth]{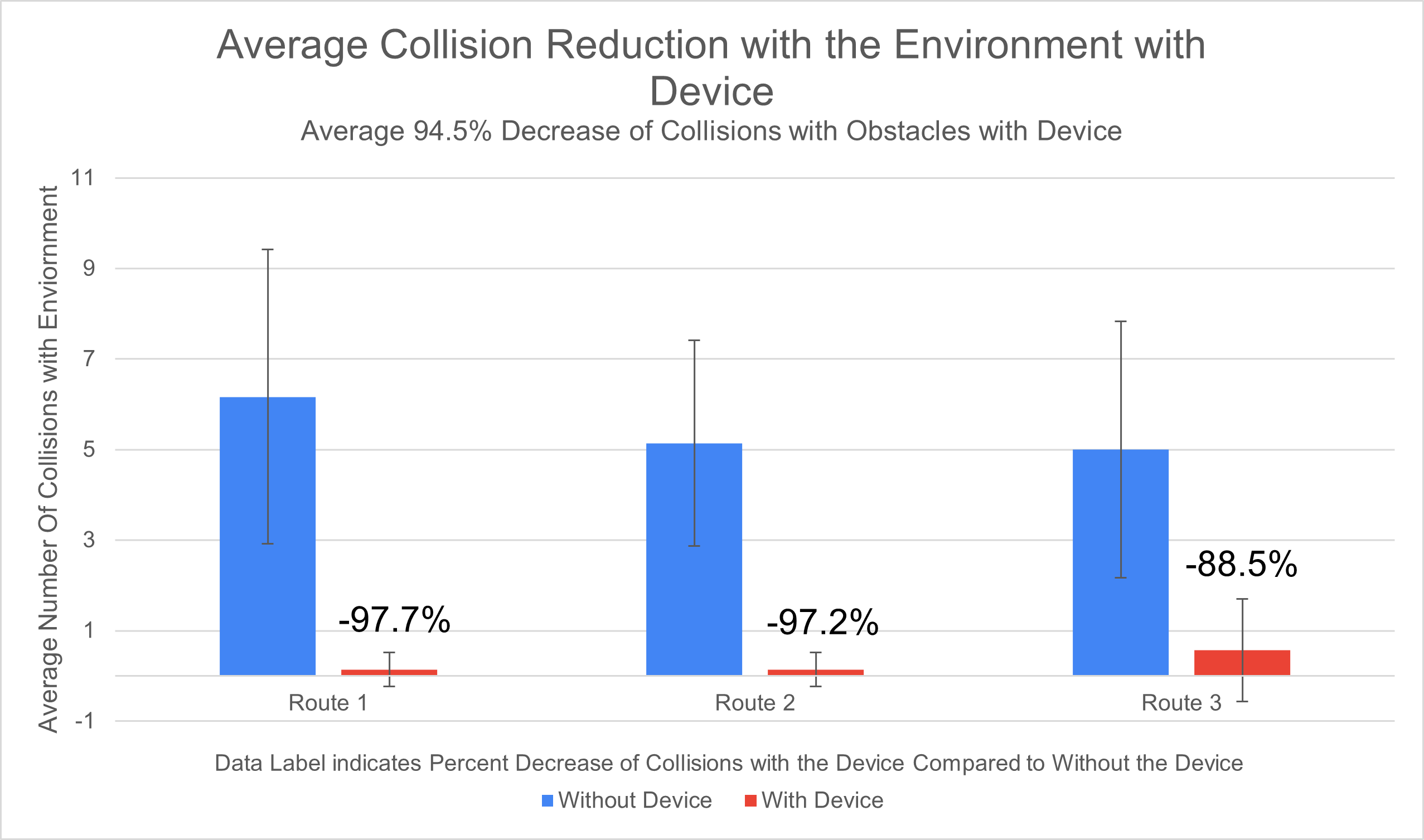}
  \caption{Memory required for Localization Model}\label{fig:figure12}
\endminipage
\end{center}
\end{figure}

\begin{figure}[!htb]
\begin{center}
\minipage{0.6\textwidth}
  \includegraphics[width=\linewidth]{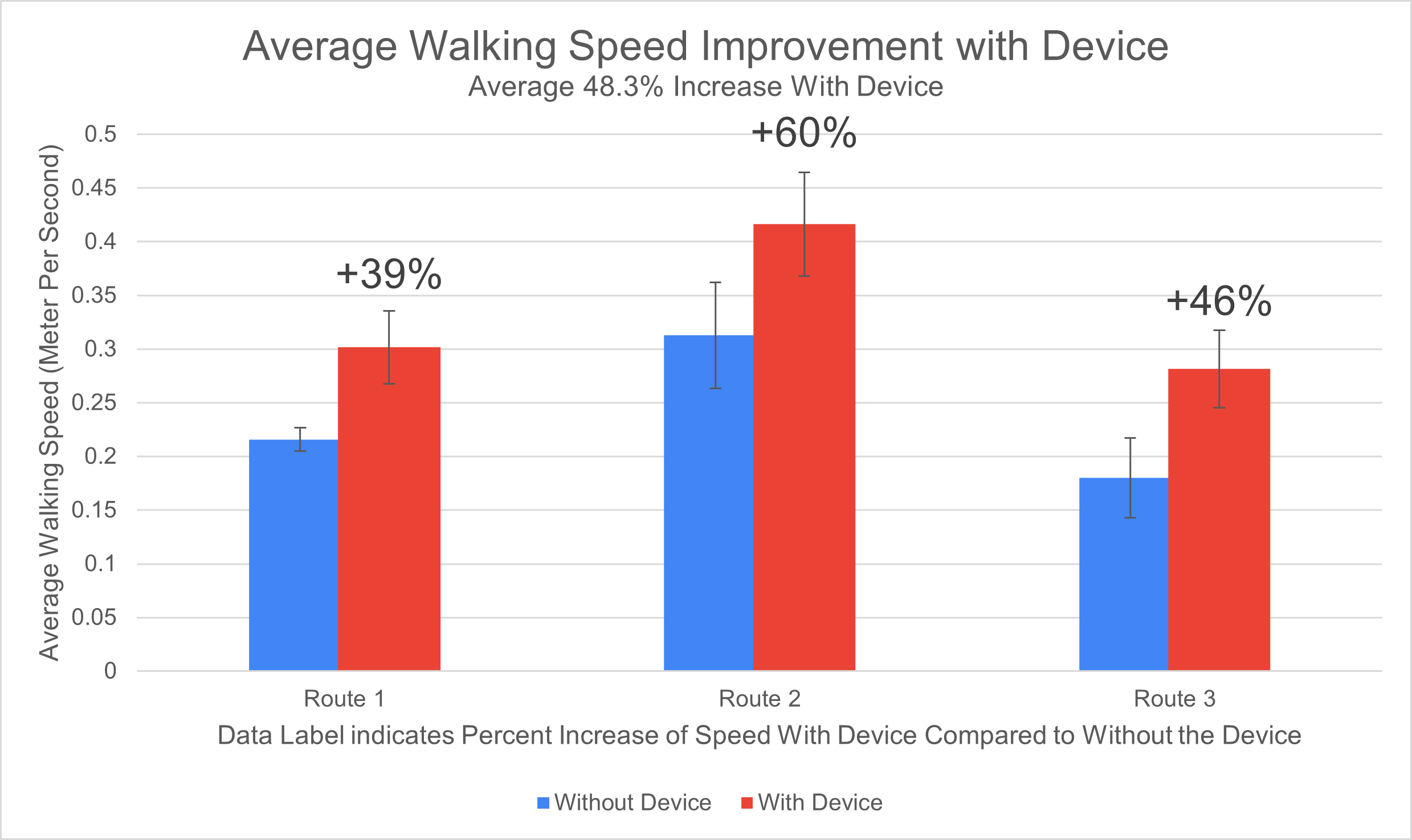}
  \caption{Memory required for Localization Model}\label{fig:figure13}
\endminipage
\end{center}
\end{figure}

\section{Analysis and Discussion}

\subsection{Dead Reckoning Based Pose Localization }

As shown in the 3D trajectories, with standard dead reckoning without gait tracking, the error in the position rapidly builds to the point where the results are not usable, even when applying a Mahony Filter for AHRS (Attitude and Heading Reference System) to help filter out noise. Instead, the device reset the velocity drift after each step to avoid drift. However, normally this process is done after the data has been collected. Instead, the device continually applies a bandpass filter to accelerometer data and detect when the acceleration magnitude is below a certain threshold to determine when the foot is stationary, then calculate the velocity drift, turn the previous step and compensate for it, immediately after each step. However, it is hard to quantify the error without absolute position tracking, which is only possible with expensive and large motion capture systems, something that this study was trying to avoid with dead reckoning. In future studies, I hope to be able to quantify the improvement of the developed method with the use of an infrared ball tracking motion capture system. Nevertheless, by analyzing the trajectory visually, it can clearly be seen that the path recorded in testing (a spiral staircase and walking down a corridor) is present, showing that gait-based drift compensation is working.

\subsection{Machine Learning-based Pose Localization}

According to this research, vision transformers are a better choice due to their higher accuracy. Vision transformers are not as parameter (22 million parameters versus 9.067 million parameters) or computationally efficient (2.9 gigaflops versus 4.6 gigaflops) compared to fully connected CNN but offer vastly improved accuracy with 1.3 meters and 13.46-degree error, compared to fully connected CNN's error of 4.8 meters and 29.4 degrees errors. This means that this research reduces the error in position by 54.2\%, and reduces orientation error by 7.8\%, a massive improvement compared to the previous control method.  Even compared to the control Residual Learning CNNs, which have a similar parameter and flops requirement, vision transformers are still more performant. This is due to the self-attention feature, which enables encoding of features across an image, compared to the standard sliding filter of a CNN. Self-attention compares each pixel to each pixel, allowing for larger details to be detected. This means, on powerful devices, vision transformers are the better choice for pose localization. When devices start to become more powerful for machine learning, such with the more widespread adoption of dedicated machine learning hardware like Apple Neural Engine, vision transformers will become the de facto choice due to superior accuracy.

However, transformers still require a larger amount of data to train on. While the training comparison dataset was quite large (4000+ samples), it would be worth evaluating how performance is affected on smaller datasets.

\subsection{Object Detection}

For a 3D CNN object detection, the number of feature levels is shown to have some effect on performance. However, at 4 levels, little increase is shown over 3 levels, and comparing HDResNet18 at 4 levels and HDResNet18 at 3 levels even shows a decrease in performance (60.44 vs 59.43 mAP at 50\% intersection), showing that at least for small-sized rooms, 3 levels can accurately process all the features in a room, without overfitting.  Increasing the encoder depth does increase detection accuracy, indicating that increasing the depth may be useful for further increasing detection accuracy, without the massive memory and computation increases needed for increasing the feature level. A further metric to analyze in the future include a larger dataset with large environments and more complicated, smaller objects, as the voxelization (dividing each scene into a grid with small boxes), may suffer when dealing with smaller objects, however, the low computation requirements (being able to process multiple scenes a second), makes it well suited for mobile devices. By optimizing the number of feature levels and depth of each model, this research increased the accuracy of object detection by 60.2\% when compared to the previous state-of-the-art model, while processing data 260\% faster, and utilizing only 41.5\% of the memory.

\subsection{Human Testing}

With the developed device, human users were able to significantly walk much faster with an average increase of 48.3\% in walking speed, as well as an average reduction of collisions by 94.5\%. This shows that the developed end-to-end system using 3D route planning and voice interface can allow for faster and more safe day-to-day indoor travel. With the low cost of 20 dollars to produce, it is also affordable to most. However, while the device was effective at navigation, it still required some human setup when installing the application and configuring before first use. Setting up the project, and installing the required code could be frustrating to non technical users. To help alleviate this potential source of frustration, in future studies, I hope to be able to create an automated installer system that installs the software and configures it automatically, without need for any human input.

\section{Conclusion}

This research develops a working system that can be used by a visually impaired user with little instruction and practice to navigate through a complicated urban environment. This work developed a new machine-learning algorithm that can be used to estimate the position and orientation of a smartphone camera, without the need for a computationally intensive SLAM-based system with 3D cameras. This study also adapted a technique for more accurate dead reckoning using gait tracking to run in real-time, enabling it to be used in real-time navigation. Finally, this research optimized a 3D object detection system to detect the position and sizes of objects in a 3D environment, to allow for more optimized path planning in a complicated urban environment. This project goes on to suggest that it is feasible to replace the method of white cane navigation with a modern, 3D-imaging solution, which allows for more efficient navigation. This should help reduce the amount of “searching” a user has to do to find objects and reach an area of interest, like the exit to a room, as well as reduce the difficulty of navigating a new and unfamiliar environment for the visually impaired. With the low cost of the developed device, costing a mere 20 dollars to prototype, the developed research also ensures that anyone can afford the device. This project can also serve as a general framework for 3D-imaging visually impaired navigation aids, something that has not existed in the past. 

\section*{Acknowledgments} 
Thank you to my parents for supporting my research, Cathy Messenger for mentoring my project, as well as the rest of 2021-2022 ASR class for encouragement and moral support along the way.

\bibliographystyle{unsrt}  
\bibliography{references}  

\end{document}